\documentclass[showpacs,preprintnumbers,pra,twocolumn]{revtex4}
\usepackage{amssymb}
\usepackage{amsmath}
\usepackage{graphicx}
\usepackage{dcolumn}
\usepackage{bm}

\begin{document}

\title{The Dual Roles of Quantum Discord in a  Non-demolition Probing
Task}
\author{Chang-shui Yu$^1$}
\email{quaninformation@sina.com; ycs@dlut.edu.cn}
\author{Jia-sen Jin$^1$}
\author{Heng Fan$^2$}
\author{He-shan Song$^1$}
\affiliation{$^{1}$School of Physics and Optoelectronic Technology, Dalian University of
Technology, Dalian 116024, P. R. China\\
$^{2}$Beijing National Laboratory for Condensed Matter Physics, Institute of
Physics, Chinese Academy of Sciences, Beijing 100190, China}
\date{\today }

\begin{abstract}
We present a non-demolition quantum information processing task of probing the information
of a class of quantum state. In this task, the information is extracted by some unitary evolution with the introduced probing qubit assisted, but  the probed
quantum state (density matrix) is  undisturbed at any time and independent of the
choice of the initial probing state.
We give a sufficient and necessary condition on the Hamiltonian that can lead to the successful realization of
such a task. We prove that,  for any feasible scheme, the probed plus probing
system will always stay at a  disentangled state
with one side quantum discord absent and the other side one inevitably produced in the probing
process. An explicit example is given for the demonstration, whilst the example shows
that the ratio of quantum discord to the total correlation will have to
reduce to zero for the maximal accessible information. In this sense, we
say that quantum discord plays the dual roles in this case.
\end{abstract}

\maketitle

\section{Introduction}

As one kind of quantum correlation, quantum
entanglement has played an important role in quantum information processing
tasks (QIPT) and is an important physical resource [1]. However, quantum
entanglement is the only quantum correlation for quantum pure states and it
can not generally cover all the quantumness of correlation for mixed states
[2-5]. Strong evidence has shown that some QIPT displayed the quantum
advantage without any entanglement in the tasks
[6-8]. On the contrary, in these QIPT without entanglement there could
be quantum discord which  is introduced to effectively distinguish the
quantum from the classical correlations [2-4].

Quantum discord `includes' quantum entanglement but beyond it due to the
potential presence in separable states [9-11]. In recent years, quantum
discord attracts increasing interests in various areas including the
dynamical evolution [11-15], the operational interpretation [16-22], the
quantification [23-25] and so on. However, the relevant research on the
roles of quantum discord played in QIPT is only restricted to several
limited cases such as the deterministic quantum computation with one qubit
(DQC1) [6] and the assisted optimal state discrimination [26] and so on
[27-29].  In particular, most of the cases show that quantum discord plays the constructive role.
Therefore, it is natural to raise the following questions. Can we find some other QITP
where it is the quantum discord instead of quantum entanglement that plays
the constructive role?  Does the quantum discord always play the
constructive role?

In this paper, we answer the questions by considering a non-demolition probing QIPT. That is,
we use a probing qubit to probe the information of a quantum state
subject to some unitary evolution, but density matrix of the probed quantum
state can not be changed throughout the probing process, no matter what kind
of initial probing state is used. We give a sufficient and necessary
condition on the Hamiltonian for the successful realization of the task. We find that no matter
how the task is implemented, the probed and probing systems will always stay
at a classical-quantum (sometimes classical-classical) disentangled state [18]. In particular,
 by a very novel approach to revealing the hidden quantum property, we find that it is
inevitable that (one-side) quantum discord will be produced in the probing process, even
though it could vanish at some special times. This shows the
constructive role of quantum discord played in the process. We also give an
explicit example to demonstrate the probing. It is interesting to find, in this example, that with the increasing of the accessible information to
the maximum, the ratio of the quantum discord to the total correlation will
decrease monotonically to zero. This means that quantum discord prevents the
effective usage of the total correlation, so it plays the negative role in
this sense. Thus we say that quantum discord in this QIPT plays the dual roles.This paper is organized as follows. In Sec. II, we 
present our non-demolition probing task and discuss the duel roles of quantum correlation. In Sec. III, we draw the conclusion.

\section{The task and the roles of quantum correlation}
\subsection{The task}
To begin with, we would like to first give a brief
description of the probing task by a class of  quantum states of qubit.
Suppose a quantum state of a qubit given by 
\begin{equation}
\rho _{u}=x\left\vert +\right\rangle \left\langle +\right\vert +\frac{1-x}{2}%
\mathbf{1}\text{,}
\end{equation}%
where $0\leq x<1$, $\left\vert \pm \right\rangle =\frac{1}{\sqrt{2}}\left(
\left\vert 0\right\rangle \pm \left\vert 1\right\rangle \right) $ and $%
\mathbf{1}$ is the identity. (i) \textit{Now we hope to extract the
information of }$x$\textit{\ such that the probed state has still the
density matrix }$\rho _{u}$\textit{\ during the probing process.} Generally,
since we have known the eigenvectors of $\rho _{u}$, we can construct an
observable $A$ made up of the projectors $\left\vert \pm \right\rangle
\left\langle \pm \right\vert $ and directly measure the ensemble $\rho _{w}$
to reach our aim. However, for some particular purpose, it is supposed that
the direct measurements on the qubit are not allowed. (ii) \textit{So we
require to introduce a probing qubit to interact with the probed one
governed by some unitary evolution.}

\subsection{The constructive role}

With the task in mind, we will proceed by
the following rigid way.

\textbf{Theorem 1.}-Let $\rho _{p}$ be the initial state of the probing
qubit and $U(t)=\exp \left( -iHt\right) $ be the unitary operation that
successfully probes $\rho _{u}$ given in Eq. (1), then the interaction
Hamiltonian $H$ can be written as 
\begin{equation}
H=\sigma _{x}\otimes A+\mathbf{1}\otimes B,
\end{equation}%
where $\sigma _{x,y,z}$ denote the Pauli matrices, $A$ and $B$ are Hermitian
with Tr$B=0$\ and $\left[ A,\rho _{p}\right] \neq 0$.

\textbf{Proof.} The final state after the evolution of the infinitesimal $%
\delta t$ governed by $U(\delta t)=\exp \left( -iH\delta t\right) $ is given
by%
\begin{equation}
\rho _{f}=\left( \mathbf{1}-iH\delta t\right) \left( \rho _{u}\otimes \rho
_{p}\right) \left( \mathbf{1}+iH\delta t\right) .
\end{equation}%
To the first order of $\delta t,$ we have 
\begin{equation}
\Delta =\left\vert \text{Tr}_{p}\rho _{f}-\rho _{u}\right\vert ^{2}=\delta
t\left\vert \text{Tr}_{p}\left[ H,\rho _{u}\otimes \rho _{p}\right]
\right\vert ^{2}
\end{equation}%
with $\left[ \cdot ,\cdot \right] $ denoting the commutator. Since any
traceless Hermitian matrix $H$ can be rewritten as%
\begin{equation}
H=\sum\limits_{j}r_{j}\left( \sigma _{j}\otimes \mathbf{1}\right)
+\sum\limits_{k}s_{k}\left( \mathbf{1}\otimes \sigma _{k}\right)
+\sum\limits_{j,k}T_{jk}\sigma _{j}\otimes \sigma _{k},
\end{equation}%
Eq. (4) can be given by 
\begin{equation}
\Delta =\delta t\sum\limits_{j}r_{j}\left[ \sigma _{j},\rho _{u}\right]
+\delta t\sum\limits_{j,k}T_{jk}\left[ \sigma _{j},\rho _{u}\right] \left( 
\text{Tr}_{p}\rho _{p}\sigma _{k}\right) .
\end{equation}%
We require that the reduced density matrix of $\rho _{f}$ is not disturbed
by any probing state $\rho _{p}$, i.e., $\Delta =0$, so 
\begin{equation}
\Delta =0\iff \left[ \sigma _{j},\rho _{u}\right] =0,
\end{equation}%
which implies that, for any $k=x,y,z$ ,
\begin{equation}
\left\{ 
\begin{array}{c}
r_{y}=r_{z}=0 \\ 
T_{yk}=T_{zk}=0%
\end{array}%
\right. .
\end{equation}%
In addition, substitute Eq. (5) into Eq. (3) and trace over the probed qubit 
$\rho _{u}$, we can arrive at 
\begin{equation}
\text{Tr}_{u}\rho _{f}=\rho _{p}-i\delta t\sum\limits_{k}s_{k}\left[ \sigma
_{k},\rho _{p}\right] -i\delta t\sum\limits_{k}T_{xk}\left[ \sigma _{k},\rho
_{p}\right] \left( \text{Tr}_{u}\rho _{u}\sigma _{x}\right) .
\end{equation}%
Insert Eq. (1) into Eq. (9), one can find that the unknown parameter $x$ is
only included by Tr$_{u}\rho _{u}\sigma _{x}$. Therefore, in order to be
able to extract the information on $x$, we require that 
\begin{equation}
\sum\limits_{k}T_{xk}\left[ \sigma _{k},\rho _{p}\right] \neq 0.
\end{equation}%
Thus $H$ can be written as 
\begin{eqnarray}
H &=&r_{x}\sigma _{x}\otimes \mathbf{1}+\sum\limits_{k}T_{xk}\sigma
_{x}\otimes \sigma _{k}+\sum\limits_{k}s_{k}\left( \mathbf{1}\otimes \sigma
_{k}\right)  \notag \\
&=&\sigma _{x}\otimes A+\mathbf{1}\otimes B,
\end{eqnarray}%
where 
\begin{equation}
A=r_{x}\mathbf{1}+\sum\limits_{k}T_{xk}\sigma _{k}
\end{equation}%
is Hermitian and 
\begin{equation}
B=\sum\limits_{k}s_{k}\sigma _{k}
\end{equation}%
is traceless and Hermitian. Meanwhile, Eq. (10) is equivalent to $\left[
A,\rho _{p}\right] \neq 0$. The proof is completed.\hfill$\blacksquare$

\textbf{Theorem 2.}-If an interaction Hamiltonian $H$ satisfies Theorem 1,
any probing process governed by $H$ will need no entanglement, but
inevitably produce (only one-side) quantum discord in the infinitesimal time.

\textbf{Proof.}-The final state after the infinitesimal evolution is given
by Eq. (3). Substitute Eq. (2) into Eq. (3), we have 
\begin{eqnarray}
\rho _{f} &=&\rho _{u}\otimes \rho _{p}-i\delta t\left[ \sigma _{x}\otimes A+%
\mathbf{1}\otimes B,\rho _{u}\otimes \rho _{p}\right]  \notag \\
&=&\frac{1+x}{2}\left\vert +\right\rangle \left\langle +\right\vert \otimes
\left( \rho _{1}-X\right)  \notag \\
&&+\frac{1-x}{2}\left\vert -\right\rangle \left\langle -\right\vert \otimes
\left( \rho _{1}+X\right)
\end{eqnarray}%
with%
\begin{eqnarray}
\rho _{1} &=&\rho _{p}-i\delta t\left[ B,\rho _{p}\right] ,  \notag \\
X &=&i\delta t\left[ A,\rho _{p}\right] .
\end{eqnarray}%
It is obvious that there is not any entanglement in $\rho _{f}$.

It is also easy to find that, if $\rho _{f}$ has no discord, $\left[ \rho
_{1},X\right] =0$. Next, we will show that this is not satisfied at any rate.
Expand $A$, $B$ and $\rho _{p}$ in the basis of the Pauli matrices, we will
have Eqs. (12,13) for $A$ and $B$ and%
\begin{equation}
\rho _{p}=\mathbf{1}+\sum\limits_{k}p_{k}\sigma _{k}.
\end{equation}%
Let $\mathbf{A}=[T_{xx},T_{xy},T_{xz}]^{T}$, $\mathbf{B}%
=[s_{x},s_{y},s_{z}]^{T}$ and $\mathbf{C=}[p_{x},p_{y},p_{z}]^{T}$ denote
vectors, one will easily check that $\left[ A,\rho _{p}\right] =\mathbf{A}%
\times \mathbf{C}$ with "$\times "$ denoting the outer product. Thus for
nonzero $\mathbf{A}$, $\mathbf{B}$ and $\mathbf{C,}$ 
\begin{eqnarray}
&&\left[ \rho _{1},X\right]  \notag \\
&=&\left[ \rho _{p}-i\delta t\left[ B,\rho _{p}\right] ,X\right]  \notag \\
&=&i\delta t\left( \mathbf{C}-i\delta t\mathbf{B}\times \mathbf{C}\right)
\times \left( \mathbf{A}\times \mathbf{C}\right)  \notag \\
&=&i\delta t\mathbf{C}\times \left( \mathbf{A}\times \mathbf{C}\right) 
\notag \\
&\neq &0.
\end{eqnarray}%
This shows that there is always quantum discord in the process. The proof is
completed.\hfill$\blacksquare$

Based on the above two theorems, we show that any  non-demolition
probing processing will need no entanglement, but always need quantum
discord within an infinitesimal evolution from the beginning. In this sense,
we say that at quantum discord plays the constructive role in the probing
task. Of course, this doesn't exclude that there could be some special time $%
t^{\ast }$ such that the whole system will arrive at a final state $\rho
_{f}(t^{\ast })$ with no quantum discord. If so it is, one could think that
if we operate a quantum gate $U(t^{\ast })=\exp \left( -iHt^{\ast }\right) $
on our initial state $\rho_u\otimes\rho_p$, we will be able to complete our
task without any discord. However, in the practical scenario, it is not the
case, because the operation of any quantum gate actually needs to evolve
from the beginning to some expected time. For a given Hamiltonian $H$ all
the parameters but the time $t$ are fixed, so for such a fixed 1-parameter
unitary group $U(t)$ the given $H$ is the unique possible Hermitian operator
based on Stone's Theorem [30]. Therefore, there is no other forms of quantum
evolution to implement the quantum gate $U(t^{\ast })$. So it is inevitable
to produce quantum discord in the process.

\subsection{The negative role and an example} 
In the previous section, we have
shown the constructive role (the necessary presence) of the quantum discord
in the coupling between the probing and the probed systems. Next, we will
give an evidence that the quantum discord in some probing process could play
the negative role by considering the usage of amount of the information. Since we need to quantitively study the
various correlations in the tasks, we will consider a concrete model of our
quantum probing instead of in a general way like the previous two theorems.
Meanwhile, this model will also give a detailed demonstration of the
previous theorems.

Suppose that the probed quantum state is still given by Eq. (1) and the
intial probing state is $\rho _{p}=\left\vert 0\right\rangle \left\langle
0\right\vert $. Let the interaction Hamiltonian governing the unitary
evolution be%
\begin{equation}
H=g\sigma _{x}^{u}\sigma _{x}^{p},
\end{equation}%
where the superscript `u' corresponds to the probed qubit and `p'
corresponds to the probing one. It is obvious that the current case
satisfies theorem 1 and 2, so the  non-demolition probing can be
successfully implemented. In addition, this Hamiltonian $H$ given in Eq.
(18) can be easily realized in spin chains systems without magnetic field
[31], Josephson junction [32] or the system of the interaction between
cavity and atoms [33]. After the evolution, one has the final state of the
joint system as%
\begin{gather}
\rho _{f}(t)=\exp \left( -ig\sigma _{x}^{u}\sigma _{x}^{p}t\right) \left(
\rho _{u}\otimes \left\vert 0\right\rangle \left\langle 0\right\vert \right)
\exp \left( ig\sigma _{x}^{u}\sigma _{x}^{p}t\right)  \notag \\
=\frac{1+x}{2}\left\vert +\right\rangle _{u}\left\langle +\right\vert
\otimes \rho _{p+}+\frac{1-x}{2}\left\vert -\right\rangle _{u}\left\langle
-\right\vert \otimes \rho _{p-}
\end{gather}%
with 
\begin{equation}
\rho _{p\pm }=\left( 
\begin{array}{cc}
\cos ^{2}(gt) & \pm i\sin (gt)\cos (gt) \\ 
\mp i\sin (gt)\cos (gt) & \sin ^{2}(gt)%
\end{array}%
\right) .
\end{equation}%
It is obvious that (I) the reduced density matrix of the probed quantum
system is $\rho _{u}$, the reduced density matrix of the probing system is 
\begin{equation}
\rho _{pf}=\left( 
\begin{array}{cc}
\cos ^{2}(gt) & ix\sin (gt)\cos (gt) \\ 
-ix\sin (gt)\cos (gt) & \sin ^{2}(gt)%
\end{array}%
\right) ,
\end{equation}%
(II) there isn't any entanglement in $\rho _{f}(t)$, and (III) $x$ can be extracted, for example,  measure $%
\sigma _{y}$ on $\rho _{f}$ at any a proper $t^{\ast }$ such that $\sin
(2gt^{\ast })\neq 0$, we will easily be able to read the value $x$ by $-%
\frac{\text{Tr(}\sigma _{y}\rho _{f}\text{)}}{\sin (2gt^{\ast })}$. (I-III) are consistent with
our previous theorems and the independence of the initial states can be found by checking that the
reduced density Tr$_{p}\rho _{f}(t)$ will not be disturbed if we use any
other quantum probing state. In addition, One can easily find that the
geometric quantum discord $D$ defined in Ref. [25] can be analytically given
by%
\begin{equation}
D\left( \rho _{f}(t)\right) =\frac{x^{2}+1-\sqrt{\tau ^{2}\left(
x^{2}-1\right) ^{2}+4x^{2}}}{8},
\end{equation}%
where $\tau =\cos \left( 4gt\right) $. It is obvious that $D\left( \rho
_{f}(t)\right) $ vanishes for $\tau =\pm 1$, i.e., $t=\frac{k\pi }{4g}%
,k=0,1,2,\cdots $. This gives an explicit demonstration that quantum discord
plays the role even though it could vanish in the evolution.

Why can one extract the information of the probed state by the measurement
on the probing state? The necessary condition for this is that there
should exist the common information (the mutual information or the total
correlations) between the probing and the probed systems due to their
interactions. However, it is not sufficient for this task. Not all the
mutual information are accessible, because the quantum part of the mutual
information has to be destroyed in any local measurements and the extraction
of the classical part is up to the employed measurements. Suppose we can
always find the optimal measurements that attain the maximal accessible
information, i.e., the classical correlation defined between the probed and
the probing systems. So on the one hand, the larger ratio of the classical
to the total correlation (usage rate) will mean the better usage of the
limited resource (total correlation). In this sense, consider the quantum discord defined as the difference between the total
and the classical correlations, the larger ratio of quantum correlation
(quantum discord) to the total correlation naturally means the lower
efficiency of the usage of the total correlation. On the other hand, the
larger classical correlation will also mean the more information to be
extracted. Next, we will show that, with the maximal accessible information
or classical correlation monotonically increasing to its maximum, the usage
rate of the total correlation will also monotonically increase to one. On
the contrary, the ratio of quantum discord to the total correlation will
monotonically decrease to zero. So we think that the quantum discord in this
case plays  the negative role. 
\begin{figure}[tbp]
\includegraphics[width=0.8\columnwidth]{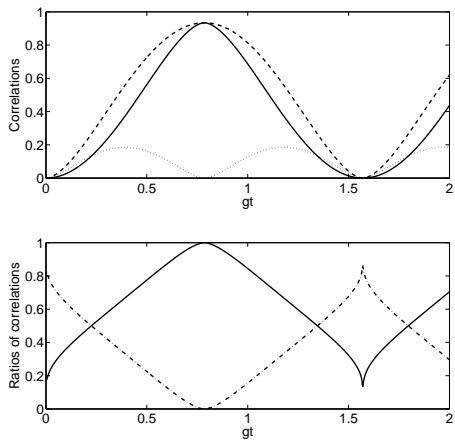}
\caption{(Dimensionless) The upper figure shows the total correlation (dashed line), the
classical correlation (solid line) and the quantum discord (dotted line) $vs.
$ $gt$. It is shown that the maximal classical correlation corresponds to
the zero quantum discord. The lower figure shows the ratios $Q\left( \protect%
\rho _{f}\right) /I\left( \protect\rho _{f}\right) $ (solid line) and $C\left( \protect%
\rho _{f}\right) /I\left( \protect\rho _{f}\right) $ (dash-dotted line) $vs.$ $gt$. The larger
ratio of classical correlation and the larger accessible information correspond to the smaller ratio of quantum
discord, which implies that quantum discord prevents the effective usage of
the total correlations by comparing the two figures.}
\label{1}
\end{figure}

To do so, we employ the the information theoretic quantum discord $Q$ and
classical correlation $C$ [4], respectively, defined for $\rho _{f}(t)$ as 
\begin{equation}
Q\left( \rho _{f}\right) =I\left( \rho _{f}\right) -C\left( \rho _{f}\right)
,
\end{equation}%
where%
\begin{equation}
I\left( \rho _{f}\right) =S(\rho _{pf})+S(\rho _{u})-S\left( \rho _{f}\right)
\end{equation}%
is the total correlation, 
\begin{equation}
C\left( \rho _{f}\right) =S(\rho _{u})-\min_{\left\{ Q_{pk}\right\}
}\sum\limits_{k}q_{k}S\left( \rho _{k}\right)
\end{equation}%
with $S(\rho )=-Tr\rho \log _{2}\rho $ the von Neumann entropy, $\rho _{k}=%
\frac{\left( \mathbf{1}\otimes Q_{pk}\right) \rho _{f}\left( \mathbf{1}%
\otimes Q_{pk}\right) }{q_{k}}$, $q_{k}=$Tr$\left( \mathbf{1}\otimes
Q_{pk}\right) \rho _{f}\left( \mathbf{1}\otimes Q_{pk}\right) $ and $%
Q_{pk}=\left\vert \psi _{k}\right\rangle \left\langle \psi _{k}\right\vert $
the projectors on the probing qubit. Lack of the analytic expressions for
the quantum and classical correlations, in Fig. 1 we plot the total, the
classical and the quantum correlations, and the ratios $Q\left( \rho
_{f}\right) /I\left( \rho _{f}\right) $ and $C\left( \rho _{f}\right)
/I\left( \rho _{f}\right) $, which is consistent with our conclusion.

\subsection{Generalization of the probing task}
Before the end, we would like
to generalize our probing task to a general case. Now, we assume the probed
state defined in $N-$dimensional Hilbert space is given by%
\begin{equation}
\rho _{u}=\sum\limits_{i}p_{i}(x)\left\vert \nu _{i}\right\rangle
\left\langle \nu _{i}\right\vert
\end{equation}%
with $p_{i}(x)\geq 0$, $\sum p_{i}(x)=1$, are some known functions on $x$
and $\left\langle \nu _{i}\right\vert \left. \nu _{j}\right\rangle =\delta
_{ij}$, and the probing state is still a qubit. The other requirements on
the task are the same as those on the qubit case. Based on a similar proof
as Theorems 1 and 2, one will obtain the following theorem.

\textbf{Theorem 3}.-Theorem 1 and Theorem 2 hold for $\rho _{u}$ given in
Eq. (1) if $H$ given in Eq. (2) is replaced by 
\begin{equation}
H=\tilde{U}G_{T}\tilde{U}^{\dagger }\otimes A+\mathbf{1}\otimes B,
\end{equation}%
where $G_{T}$ is one generator of the $SU(N)$ group or $G_{T}=%
\sum_{i=1}^{k}a_{i}G_{i},k>1,$ with $a_{i}$ real and $G_{i}$ being some
commutation generators of $SU(N)$ group, and $\tilde{U}$ is the unitary
transformation that changes $\left\{ \left\vert \nu _{i}\right\rangle
\right\} $ to the eigenvectors set of $G_{T}$. In addition, $A$ and $B$ are
Hermitian with Tr$B=0$\ and $\left[ A,\rho _{p}\right] \neq 0$.

\section{Conclusion and discussions}
In summary, we have presented a task to
probe a single-parameter quantum state. In this scheme, the probed state is
not disturbed no matter what kind of initial probing state is chosen, so it
is non-demolition and independent of the choice of the initial probing state
(actually it is implied that the external noise on the probing qubit can not
disturb the probed state either). It can be found that that there is no
entanglement in the whole probing process and the whole bipartite system
always stays at a classical-quantum state. It is quite interesting that it
is inevitable that quantum discord will be produced for the successful probing,
but the fraction of the quantum discord in the total correlation will have to
reduce to zero for the maximal accessible information. In this sense, we say
that quantum discord in this task plays the dual roles. In addition, the
approach employed to consider the role of quantum discord could shed new
light on the understanding of quantum correlation in other QIPTs.

\section{Aknowledgement}
This work was supported by the National Natural Science Foundation of China,
under Grant No. 11175033 and `973' program No. 2010CB922904 and the
Fundamental Research Funds of the Central Universities, under Grant No.
DUT12LK42.

\end{document}